\documentclass[useAMS,usenatbib]{mn2e}
\usepackage{psfig, epsf, epsfig}
\title[Dynamical friction in galaxies]{Dynamical friction
of star clusters against disk field stars in   galaxies: Implications on 
stellar nucleus formation and globular cluster luminosity functions}
\author[K. Bekki]
{Kenji Bekki${}^1$\thanks{E-mail:
bekki@phys.unsw.edu.au}\\
       ${}^1$School of Physics, University of New South Wales,
              Sydney, NSW,  2052,  Australia\\}

\begin{document}

\date{Accepted, Received 2005 February 20; in original form }

\pagerange{\pageref{firstpage}--\pageref{lastpage}} \pubyear{2005}

\maketitle

\label{firstpage}

\begin{abstract}

We numerically investigate orbital evolution of star clusters 
(SCs) under the  influence of dynamical friction by field stars of
their host disk galaxies embedded in dark matter halos.
We find that 
SCs with masses larger than $\sim 2 \times
10^5 {\rm M}_{\odot}$ can show significant orbital decay 
within less than 1 Gyr due to
dynamical friction by disk field stars in galaxies
with disk masses ($M_{\rm d}$) less than $10^9 {\rm M}_{\odot}$.  
We also find that orbital decay of SCs due to dynamical friction
is more remarkable in disk galaxies  with smaller $M_{\rm d}$
and higher mass-ratios of disks to dark matter
halos.
The half-number radii ($R_{\rm h, sc}$) and mean masses within $R_{\rm
h, sc}$ of the SC systems (SCSs)  in low-mass disk galaxies 
with $M_{\rm d} \le  10^9 {\rm M}_{\odot}$
are found to evolve significantly with time 
owing to dynamical friction of SCs.
More massive SCs that can spiral-in to the central regions of disks
can form multiple SC systems with smaller velocity dispersions
so that they can merge with one another to form single stellar
nuclei with their masses comparable to $\sim 0.4$\% of their host disk 
masses.
Based on these results, we suggest that luminosity functions (LFs)  for more
massive globular clusters (GCs) with masses larger than
$2\times 10^5 {\rm M}_{\odot}$   can steepen
owing to transformation of the more massive  GCs into single stellar
nuclei through GC merging in less luminous  galaxies.
We also suggest that the half-number radii of GC systems can evolve
owing to dynamical friction only for galaxies with their total masses
smaller than $\sim 10^{10} {\rm M}_{\odot}$.

\end{abstract}

\begin{keywords}
globular clusters:general -- 
galaxies:formation -- 
galaxies:kinematics and dynamics -- galaxies:halos -- galaxies:star
clusters
\end{keywords}

\section{Introduction}

Dynamical friction of SCs
in galaxies have been discussed in many
different contexts, such as formation of stellar galactic nuclei
via merging of old GCs (e.g., Tremaine et al. 1975),
transformation from non-nucleated dwarfs into nucleated ones
(e.g., Oh \& Lin 2000),
physical meanings for the presence of the GC system (GCS) in
the Fornax dwarf galaxy (e.g., Oh et al. 2001),
and dark matter distributions of dwarf galaxies
(e.g., Hernandez \& Gilmore 1998; Goerdt et al. 2006; Inoue 2009).
Orbital decay of GCs due to dynamical friction can
significantly change spatial distribution of GCs and thus
increase possibilities of them to be destroyed by strong galactic
tidal fields (e.g., Vesperini 2000).
Dynamical friction of SCs (including GCs) in galaxies
is thus considered to be important for better
understanding the evolution of GC luminosity
functions (GCLFs) and that of half-number radii of the
GCSs  in galaxies
(e.g., Vesperini 2000).
Recently, van de Ven \& Chang (2009) have suggested the importance of
dynamical friction in the dynamical evolution of SCs around nuclear rings in galaxies. 

\begin{table*}
\centering
\begin{minipage}{185mm}
\caption{The ranges of model parameters.}
\begin{tabular}{cccccccc}
Parameters &
{$M_{\rm gal}$ ($\times 10^{10} M_{\odot}$)
\footnote{The total mass of a galaxy. }} &
{$M_{\rm d}$ ($\times 10^{9} M_{\odot}$)
\footnote{The total mass of a disk. }} &
{$f_{\rm d}$
\footnote{The mass fraction of stellar disk 
($=M_{\rm d}/M_{\rm gal}$) in a galaxy.}}& 
{$f_{\rm b}$ 
\footnote{The mass fraction of a bulge ($=M_{\rm b}/M_{\rm d}$) in a galaxy .}}&
{Galaxy type
\footnote{HSB and LSB represent low-surface brightness and high-surface
brightness stellar disks, respectively. }}&
{$M_{\rm V, low}$ 
\footnote{The 
lower luminosity cut-off  in the SC luminosity function.
This parameter is fixed at  $-6$ mag in all models (see the main text for main
reasons for this).}}&
{$M_{\rm V, high}$ 
\footnote{The 
higher luminosity cut-off in the SC luminosity function.}} \\
Value ranges & 0.1--10.0 & 0.1--10.0 & 0.01--0.1  & 0.0--1.0  & LSB or HSB &
$-6$ (mag) & $-12 \sim -8$ (mag) \\
\end{tabular}
\end{minipage}
\end{table*}

Dynamical friction processes of SCs
against  {\it disk field stars in disk galaxies} are important
for the following three reasons.
Firstly, recent cosmological hydrodynamical simulations on
the possible formation sites of first GCs 
(Kravtsov \& Gnedin 2005) have shown that
the present-day GCs can be formed within disk galaxies
at high redshifts ($z\sim 3$).
Secondly,  less luminous disk galaxies like the Large
Magellanic Cloud (LMC) have disky GC systems in which GCs
have disky spatial distributions and rotational kinematics
(e.g., Freeman et al. 1983; Olsen et al. 2004).
Thirdly, previous theoretical studies suggested that
the present-day dwarf ellipticals (dEs) were previously
less luminous disk galaxies with no/little bulges (Mastropietro et al.
2004).
Furthermore,
previous observational results on physical properties
of dwarf galaxies (e.g., Stiavelli et al. 2001) suggest similarity
in nuclear density profiles of stars  between dEs and spirals with
exponential bulges: these appear to imply that understanding 
nucleus formation processes 
due to SC migration caused by dynamical friction in spirals
can further lead to better understanding those in dwarfs.
Although dynamical friction processes of SCs
{\it against dark matter halos}  have been
well investigated by previous works (e.g., Goerdt et al. 2006),
those by {\it field stars in disk galaxies} have not been done
by self-consistent numerical simulations  so far.

The purpose of this paper is thus to investigate
dynamical friction of SCs against  disk field stars based on fully
self-consistent numerical simulations on 
dynamical evolution of  disk galaxies
with SCs.
We particularly investigate how dynamical friction of SCs against
disk field stars depend on physical properties of their host
galaxies, such as disk masses,  bulge-to-disk-ratios,  and  mass-ratios
of disk to dark matter halo.
Based on the present numerical results, 
we discuss the observed luminosity functions of GCs
dependent on luminosities of their host galaxies
(e.g., Jord\'an et al. 2006),  formation processes of stellar galactic nuclei,
and evolution of physical properties of GCSs in galaxies.

The plan of the paper is as follows: In the next section,
we describe our  numerical models  for orbital evolution of SCs
in disk galaxies. 
In \S 3, we
present the numerical results
mainly on (i)  orbital evolution of SCs 
and (ii) the physical properties of SCSs 
for variously different models.
In \S 4, we discuss wide implications of the present results
such as formation of stellar galactic nuclei by SC merging
and evolution of mass and luminosity functions of SCs due to
orbital decay of SCs caused by dynamical friction.
We summarize our  conclusions in \S 5.

\section{The model}

We investigate orbital evolution of SCs in disk galaxies
embedded in massive dark matter halos using  
the latest version of GRAPE
(GRavity PipE, GRAPE-7) which is the special-purpose
computer for gravitational dynamics (Sugimoto et al. 1990).
All SCs are represented by point-mass particles rather than
self-gravitating N-body systems so that we can not discuss
internal stellar dynamics of SCs influenced by galactic potentials.
Although evolution of SCs 
(e.g., internal stellar dynamics) from their formation within GMCs in a low-mass
disk galaxy is discussed by our previous numerical simulations
(e.g., Hurley \& Bekki 2008), the present numerical code
does not allow us to discuss this important issue:
we plan to discuss this in our forthcoming papers.

Galaxies are represented by N-body particles so that SCs can feel
gravitational influences of the {\it live} galactic potentials
(e.g., dynamical friction).
Non-axisymmetric structures such as bars
and spiral arms in disk galaxies  can change  background   stellar  distribution
and kinematics, which can be important for the effectiveness
of dynamical friction that depends on density and velocities
of background stars. 
Therefore,
dynamical friction of SCs against  disk field stars 
can be very complicated in the present study.
Dynamical friction of disk SCs 
(i.e., SCs initially in stellar disks) 
against disk field stars is much more
effective than  that against dark matter halos in the present study.
Therefore, the orbital decay of disk SCs demonstrated
in the present study  is due mainly
to gravitational interaction between disk field stars and SCs.

\subsection{Disk galaxy model}

Since our numerical methods for modeling dynamical evolution
of  late-type disk galaxies have already been described
by Bekki \& Peng (2006), we give only a brief
review here.
The total disk mass and the size of a disk of a  disk galaxy
with the total mass of $M_{\rm gal}$
are $M_{\rm d}$ and $R_{\rm d}$, respectively.
Time is   measured in units of
$t_{\rm g}$ = $(R_{\rm
d}^{3}/GM_{\rm gal})^{1/2}$, where $G$ is the
gravitational constant and assumed to be 1.0.
If we adopt $M_{\rm gal}$ = 1.0 $\times$ $10^{10}$ $ \rm
M_{\odot}$ and $R_{\rm d}$ = 2.3 kpc as a fiducial value, then
$t_{\rm g}$ = 1.6
$\times$ $10^{7}$ yr.
The model with this fiducial value
is referred to as ``the standard model''.
The disk is composed of a dark matter halo,
a stellar disk, and a stellar bulge.
We mainly investigate late-type, bulge-less galaxies, because
we mainly discuss SC evolution in low-mass galaxies.

The mass ratio of the dark matter halo (with the mass
of $M_{\rm dm}$) to the stellar disk
in a disk model is a free parameter ranging from 9 to 49.0
in order to investigate galaxies with different $M_{\rm gal}$ thus
different mass-to-light-ratio. 
We adopt the density distribution of the NFW
halo (Navarro, Frenk \& White 1996) suggested from CDM simulations:
\begin{equation}
{\rho}(r)=\frac{\rho_{0}}{(r/r_{\rm s})(1+r/r_{\rm s})^2},
\end{equation}
where  $r$, $\rho_{0}$, and $r_{\rm s}$ are
the spherical radius,  the characteristic  density of a dark halo,  and the
scale
length of the halo, respectively.
The $c$ parameter ($=r_{\rm s}/r_{\rm vir}$, where $r_{\rm vir}$
is the virial radius of the NFW profile)  for a galaxy
with $M_{\rm dm}$
is chosen according to the predicted $c$-$M_{\rm dm}$ relation
in the $\Lambda$CDM simulations (e.g., Neto et al. 2007). For example,
reasonable $c$ values 
for disk galaxies with 
$M_{\rm gal}=10^{10} {\rm M}_{\odot}$
are  12.9  (also
$r_{\rm vir}$ is $\sim 10 R_{\rm d}$). 
The stellar bulge with a mass $M_{\rm b}$
and size $R_{\rm b}$ is  represented by the Hernquist
profile with the scale-length of $0.2R_{\rm b}$.
The bulge mass fraction ($M_{\rm b}/M_{\rm d}$) is referred to
as $f_{\rm b}$.

The radial ($R$) and vertical ($Z$) density profiles of the disk are
assumed to be proportional to $\exp (-R/R_{0}) $ with scale
length $R_{0}$ = 0.2 and to ${\rm sech}^2 (Z/Z_{0})$ with scale
length $Z_{0}$ = 0.04 in our units, respectively: the  stellar 
disk follows this exponential distribution.
In addition to the   
rotational velocity caused by the gravitational field of disk,
bulge, and dark halo components, the initial radial and azimuthal
velocity dispersions are assigned to the disc component according to
the epicyclic theory with Toomre's parameter $Q$ = 1.5.  The
vertical velocity dispersion at given radius is set to be 0.5
times as large as the radial velocity dispersion at that point,
as is consistent with the observed trend of the Milky Way (e.g.,
Wielen 1977).

We investigate  models with different $M_{\rm gal}$ and adopt
the Freeman's law (Freeman 1970) to determine $R_0$ 
of a disk galaxy according to its disk mass:
\begin{equation}
R_{\rm 0}=3.5 {(\frac{M_{\rm d}}{6\times 10^{10} {\rm
M}_{\odot}})}^{0.5} {\rm kpc.}
\end{equation}
Structural and kinematical properties of dark matter halos
and stellar disks are assumed to be self-similar between
models with different $M_{\rm gal}$.
We also investigate  low-surface brightness (LSB)
models with different $M_{\rm gal}$ in which 
$R_0$ 
of a disk galaxy is determined as:
\begin{equation}
R_{\rm 0}=8.8 {(\frac{M_{\rm d}}{6\times 10^{10} {\rm
M}_{\odot}})}^{0.5} {\rm kpc.}
\end{equation}

Both the total particle number of  dark matter halo ($N_{\rm h}$)
and that of the stellar disk ($N_{\rm d}$)
in a bulgeless disk model are 500000.
The total number for  bulge ($N_{\rm b}$) in a model depends on
$M_{\rm b}$ such that $N_{\rm b}=N_{\rm d}M_{\rm b}/M_{\rm d}$.
The fixed gravitational softening length is $0.008R_{\rm d}$,
which corresponds to 19pc in the standard model with
$M_{\rm d}=10^9 {\rm M}_{\odot}$. The mass-ratio of the lowest-mass 
SC to each disk field star is more than 10 owing to the adopted
particle number ($\sim 10^6$) so that we can properly investigate dynamical
friction processes of SCs by disk field stars.
The leapfrog integration scheme with a fixed time step of $0.02t_{\rm g}$ is 
adopted for all models.

\begin{figure}
\psfig{file=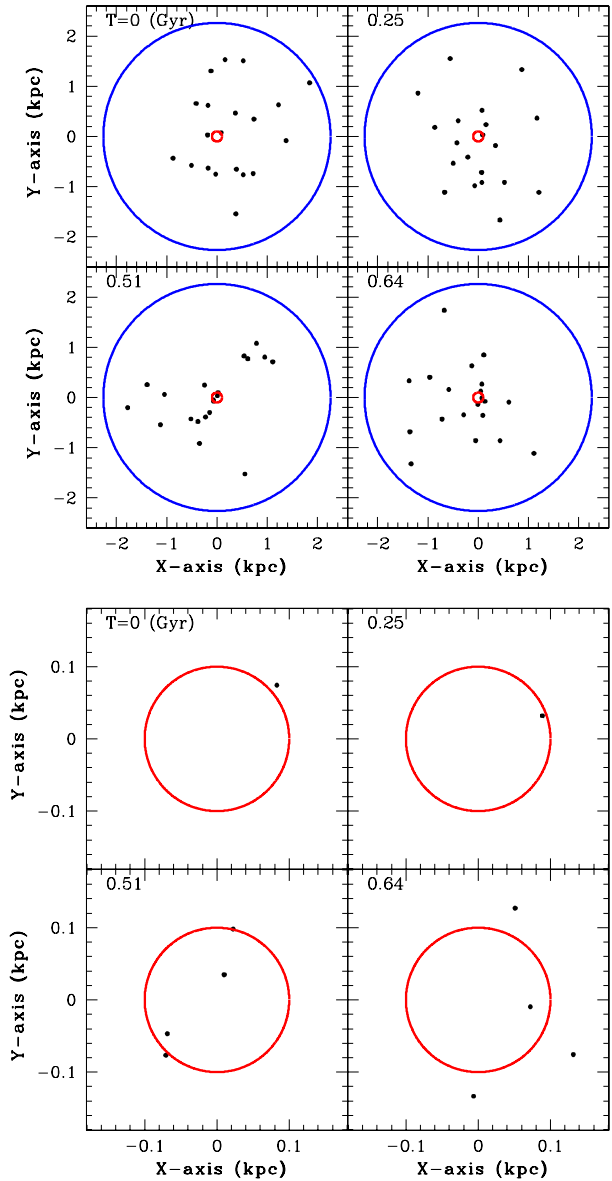,width=8.0cm}
\caption{
Spatial distributions of disk SCs represented by small black dots
in the disk galaxy projected onto the $x$-$y$ plane
for four different time steps in the standard model.
Upper and lower panels show the entire and nuclear distributions
of SCs, respectively.
Time $T$ in units of Gyr is given in each of the four panels.
Big blue and small red circles represent the initial disk size
of the galaxy and the nuclear region ($R=100$ pc), respectively.
}
\label{Figure. 1}
\end{figure}

\begin{figure}
\psfig{file=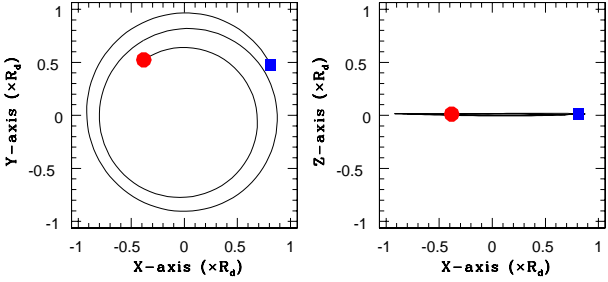,width=8.0cm}
\caption{
Orbital evolution of a SC with $m_{\rm sc}=
3.3 \times 10^5 {\rm M}_{\odot}$
projected onto the $x$-$y$ plane (left)
and the $x$-$z$ plane (right).
The initial and final locations of the SC
are indicated by a blue square and by a red circle,
respectively.
The orbit for $0\le T\le 20t_{\rm g}$ is shown
so that the orbital decay due to dynamical friction
can be more clearly seen.
}
\label{Figure. 2}
\end{figure}

\begin{figure}
\psfig{file=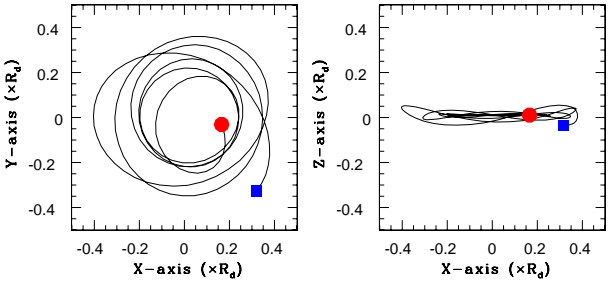,width=8.0cm}
\caption{
The same as Fig. 2 but for
$m_{\rm sc}= 1.3 \times 10^6 {\rm M}_{\odot}$.
}
\label{Figure. 3}
\end{figure}

\subsection{SC model}

Since we focus on orbital evolution of long-lived SCs in disk galaxies,
we adopt the canonical luminosity function (LF) observed for old GCs
(Harris 1991):
\begin{equation}
\Phi (M)=C \times {\rm e}^{-(M-M_0)^2/2{{\sigma}_m}^2},
\end{equation}
where $C$ is a constant,
$M$ is the magnitude of a SC, $M_0$=$-7.27$ mag 
in the $V$ band (Harris 1991),
and ${\sigma}_{\rm m}=1.2$ mag.
We assume that  $M_0$ and ${\sigma}_{\rm m}$
should be the same between different models in order to
show more clearly how dynamical friction processes of SCs
depend on physical properties of their host galaxies.
The mass function (MF) of SCs (SCMFs) is  derived from
the LF os SCs (SCLFs) for a reasonable
stellar-mass-to-light-ratio.

The $V$-band magnitudes of SCs for lower and higher
cut-off luminosities/masses  in SCLFs/SCMFs  are 
described as $M_{\rm V, low}$ and $M_{\rm V, high}$, 
respectively. 
$M_{\rm V, low}$ and $M_{\rm V, high}$ are set  to be
$-6$ mag and $-10$ mag, respectively for most models.
 Although we consider
these values to be reasonable and realistic, we
investigate models with different  $M_{\rm V, high}$
for comparison. 
If we adopt $M_{\rm V, low}$ significantly larger than $-6$ mag,
the mass of a SC can be comparable to  each disk field star
in some models so that we can not properly investigate
dynamical friction processes.
We therefore consider that it is the best for the present study
to fix $M_{\rm V, low}$  at a reasonable value.
The stellar-to-mass-to-light-ratio is 3.2 for
SCs in all models to allocate a mass  ($m_{\rm sc}$) to each
SC particle.

The total number of SCs in a galaxy is determined by $S_{\rm N}$ of the galaxy,
where $S_{\rm N}$ is the specific frequency of the SC system. 
The specific frequency is defined as follows (Harris \& van den Bergh 1981):
\begin{equation}
S_{\rm N}=N_{\rm sc} \times 10^{0.4(M_{\rm v}+15)},
\end{equation}
where $N_{\rm sc}$ and $M_{\rm v}$ are the total number of SCs
in a galaxy and $V-$band absolute magnitude of the galaxy, respectively.
Thus,  $N_{\rm sc}$ is determined by $S_{\rm N}$ and $M_{\rm d}$ with
a reasonable assumption on the  mass-to-light-ratio for stars
(for converting  $M_{\rm d}$ into $M_{\rm v}$ in galaxies).

The observed dependence of 
$S_{\rm N}$ on $M_{\rm V}$ has  ``U-shape" in the sence
that the galaxy luminosity dependence
of $S_{\rm N}$ is different between galaxies
below and above a threshold luminosity, $M_{\rm V,th}$,
which is around $-19.5$ mag (Bekki et al. 2006).
We here use the  following form for the $S_{\rm N}$-$M_{\rm V}$
relation (Bekki et al 2006);
\begin{equation}
S_{\rm N} (x) = A_1 \times 10^{K_1 x}+ (S_{\rm N,th}-A_1) \times
10^{K_2 x},
\end{equation}
where $x=\frac{M_{\rm V}-M_{\rm V,th}}{M_{\rm V,th}}$ and
$S_{\rm N,th}$ is  $S_{\rm N}$ at $M_{\rm V}=M_{\rm V,th}$.
Parameter values of $A_1$, $K_1$, $K_2$, $M_{\rm V,th}$,
and $S_{\rm N,th}$ can be determined by fitting to observations.
A reasonable model fit to observational data sets have 
$M_{\rm V,th}=-19.5$ mag,
$S_{\rm N,th}=1.0$, $A_1=0.5$,  $K_1=-6$, and  $K_2=4$
(see Fig. 1 in Bekki et al. 2006).

We mainly investigate the ``standard'' SC model in which 
$M_{\rm d}=10^9 {\rm M}_{\odot}$,
$M_{\rm d}/L =4.0$ (where $L$ is the total luminosity of the disk),
$S_{\rm N}=6.8$ (thus $N_{\rm sc}=20$). This model can be consistent
with the observed U-shaped $S_{\rm N}$-$M_{\rm V}$
relation.  The models with the same $S_{\rm N}$-$M_{\rm V}$ relation
as that adopted in the standard model have $N_{\rm sc}=3$ for
$M_{\rm d}=10^8 {\rm M}_{\odot}$
and $N_{\rm sc}=117$ for $M_{\rm d}=10^{10} {\rm M}_{\odot}$
in the present study.  We however investigate  models with
$N_{\rm sc}$ smaller/larger  than those consistent with the adopted
$S_{\rm N}$-$M_{\rm V}$ relation for comparison.

We assume that the initial distribution and kinematics
of the SCS in a disk galaxy are exactly the same as those of the disk components
(i.e., rotating, exponential disk).  This is a reasonable assumption,
given that the vast majority of stars are formed
as bound and unbound SCs in the Galactic disk (e.g., Lada \& Lada 2003).
For the adopted exponential distribution of a SCS, we allocate each SC 
particle 3D-velocities as follows. We first search  for the nearest 
disk particle (i.e., disk field star)
for each SC particle and then allocate the 3D velocities of the disk particle
to the SC one. Thus the initial SCS in a galaxy has a  thin disk configuration
and rotational kinematics. 
 
In order to discuss the effectiveness of dynamical friction of disk SCs
against  disk field stars, we also briefly investigate orbital evolution
of halo SCs that are initially in halo regions of galaxies.
The number of halo SCs is the same as that of disk SCs in a galaxy
and its spatial distribution is described as  a power-law
with the power-law index of $-3.5$ as observed for the Galactic GCs
(e.g., van den Bergh 2000).
The initial half-number radius of the halo SCs in a galaxy
is $1.4R_0$ as observed for the Galactic GCS (i.e., the half-number radius
of 5 kpc for the GCS and $R_0=3.5$  kpc for the stellar disk; van der Bergh 2000).
Kinematical properties of the halo SCs are assumed to be the same as those
of the dark matter halos of their host galaxies (i.e., isotropic velocity dispersions
determined by mass distributions of the galaxies).

\subsection{Parameter study}

We mainly investigate dynamical friction processes of SCs with
different initial masses,  their orbital evolution
and the final distribution of SCs in the standard model
with $M_{\rm d}=10^{9} {\rm M}_{\odot}$, $f_{\rm b}=0$,
$f_{\rm d}=0.1$, $N_{\rm sc}=20$, 
and $M_{\rm V, high}=-10$ mag.  
We also investigate  40 models  with different  $M_{\rm d}$,
$f_{\rm b}$, $f_{\rm d}$, $N_{\rm sc}$,
$M_{\rm V, high}$,  central densities of
stellar disk (i.e., whether disks are LSB or HSB, where
LSB and HSB represent low-surface brightness and high-surface brightness
disk galaxies, respectively)
and structures of dark matter halos in order to better understand
the effectiveness of dynamical friction in disk galaxies
with different physical properties. 
The range of model parameters investigated
are shown in the Table 1.

In order to discuss formation of stellar nuclei in disk galaxies,
we investigate time evolution of the total mass of SCs within $0.1 R_{\rm d}$
for each model. We consider that SCs transfered to the central $0.1 R_{\rm d}$
owing to dynamical friction
can quickly merge with one another to form a single massive SC that can be identified
either as a nuclear SC or a stellar galactic nucleus. 
Since SCs are represented by point-mass particles in the present simulations,
nucleus formation via SC merging can not be investigated in a fully self-consistent
manner. Previous and recent numerical simulations in which SCs are represented by
N-body particles have confirmed that nuclear SCs can merge to form a single nucleus  in a
galaxy (e.g., Capuzzo-Dolcetta \& Miocchi 2008).
 We thus consider that it is appropriate for the present study
to briefly discuss formation of stellar galactic nuclei via SC merging in central
regions of galaxies based on our simulations.

Thus the total mass of the stellar nucleus
in a disk galaxy at each time step in each simulation
($M_{\rm nuc}$) is described as follows:
\begin{eqnarray}
M_{\rm nuc}= \sum_{i=1}^{N_{\rm nuc}} m_{\rm sc, \it i},
\end{eqnarray}
where $N_{\rm nuc}$ is the total number of nuclear SCs within
$0.1 R_{\rm d}$ and $m_{\rm sc,  \it i}$ is the mass of the individual
nuclear SCs.
$M_{\rm nuc}$ is not literally the nucleus mass
of the galaxy so that it can change with time relatively rapidly.
For all models, we run models for  at least $40 t_{\rm g}$ in order to
investigate orbital decay of SCs due to dynamical friction in galaxies.
We however investigate some models,
in particular, more massive disk ones,  for longer time scales (up to $200 t_{\rm g}$)
so that we can confirm whether dynamical friction is really ineffective
in these massive disk models.

The final $M_{\rm nuc}$ in a galaxy
depends strongly on
the $S_{\rm N}$ such that $M_{\rm nuc}$ can be  larger in models
with larger $S_{\rm N}$  for a fixed $M_{\rm d}$.
Although this suggests that the observe higher $S_{\rm N}$ in
nucleated dwarfs (dE,Ns; Miller \& Lotz 2007) is due to initially high $S_{\rm N}$
in dE,Ns,
we do not discuss extensively this problem based on results of models
with different $S_{\rm N}$ (by adopting $S_{\rm N}$ different
from those used in the observed  $S_{\rm N}-M_{\rm V}$ relation):
we will discuss this problem in our forthcoming papers.
In the followings,
the  time $T$ represents the time that has elapsed
since
the simulation starts.

\subsection{Comparison with analytical works}

Previous analytical works showed that
dynamical friction processes  of SCs within galaxies
depend on masses of SCs  and physical properties of their host galaxies
(e.g. Binney \& Tremaine 1987).  
These analytical  works would be quite useful and helpful in better understanding
numerical results, though they did not investigate dynamical
friction of SCs against disk field stars.
The time-scale of dynamical friction  ($t_{\rm fric}$) for typical GCs
with masses ($M_{\rm gc}$) of $2 \times 10^5 {\rm M}_{\odot}$
within galactic halos of luminous galaxies
is considered to be quite long
(e.g., Binney \& Tremaine 1987).
For the Galaxy with the circular velocity $v_{\rm c}=220$ km s$^{-1}$
and the Coulomb logarithm $\ln \Lambda=10$,
$t_{\rm \bf fric}$ can be estimated as follows:
\begin{eqnarray}
t_{\rm  fric}= 1.2 \times 10^{11}
{ ( \frac{ r_{\rm  i} } {\rm 2  kpc} )  }^{2}
( \frac{ v_{\rm  c} } {220 {\rm  km s^{-1} }} )
{ ( \frac{ M_{\rm  gc}  } {2 \times 10^5  {\rm  M}_{\odot}   } )
}^{-1}
{\rm yr},
\end{eqnarray}
where $r_{\rm i}$ is the distance of a GC from the center of
its host galaxy.
For a galaxy with $v_{\rm c}=70$ km s$^{-1}$
and $\ln \Lambda=10$,
\begin{eqnarray}
t_{\rm  fric}= 1.8 \times 10^{9}
{ ( \frac{ r_{\rm  i} } {\rm 1  kpc} )  }^{2}
( \frac{ v_{\rm  c} } {70 {\rm  km s^{-1} }} )
{ ( \frac{ M_{\rm  gc}  } {10^6  {\rm  M}_{\odot}   } )  }^{-1}
{\rm yr}.
\end{eqnarray}
Therefore, although dynamical friction  can not be important for evolution of normal SCs and GCs
within luminous galaxies like the Galaxy,
it can significantly change physical properties of SCs and those of
SCSs in dwarf galaxies.
The above results also show that dynamical friction of SCs can be more effective
for SCs initially located in the inner regions of galaxies.

These estimation is for dynamical friction of SCs against {\it galactic halos}
so that we can not rely on  these for understanding the present numerical results
for those against {\it disk field stars}.
However, the above analytical estimation would be still useful
for (i) understanding  the differences in $t_{\rm fric}$ between
dynamical friction against galactic halos and that against disk field
stars and (ii) discussing dependences of the present
results on $M_{\rm d}$ (thus on $v_{\rm c}$) in a qualitative manner.

\begin{figure}
\psfig{file=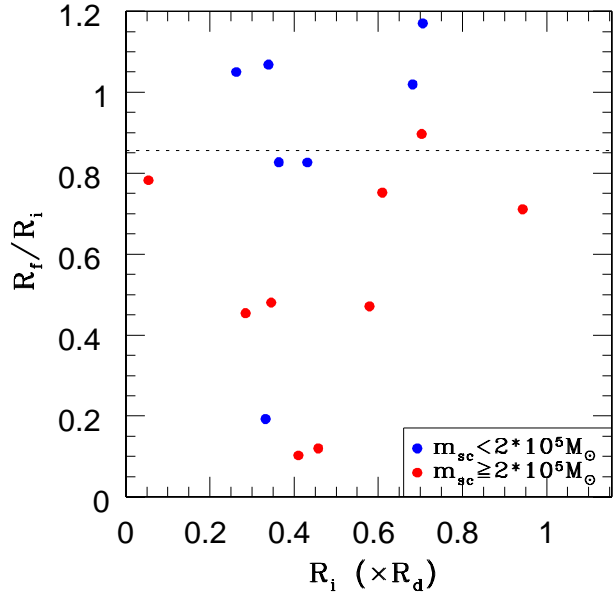,width=8.0cm}
\caption{
The ratios of final $R$ ($R_{\rm f}$) to initial one ($R_{\rm i}$)
as a function of $R_{\rm i}$ for SCs in the standard model.
Blue and red dots represent SCs with 
$m_{\rm sc} <  2 \times 10^5 {\rm M}_{\odot}$
and $m_{\rm sc} \ge  2 \times 10^5 {\rm M}_{\odot}$, respectively.
The horizontal dotted line indicates the mean value of $R_{\rm f}/R_{\rm i}$
for all SCs in this model. If we estimate the mean for SCs with
$R_{\rm f}/R_{\rm i} \le 1$ for which orbital decay due to dynamical
friction is clearly seen,  the mean value is 0.55.
}
\label{Figure. 4}
\end{figure}

\begin{figure}
\psfig{file=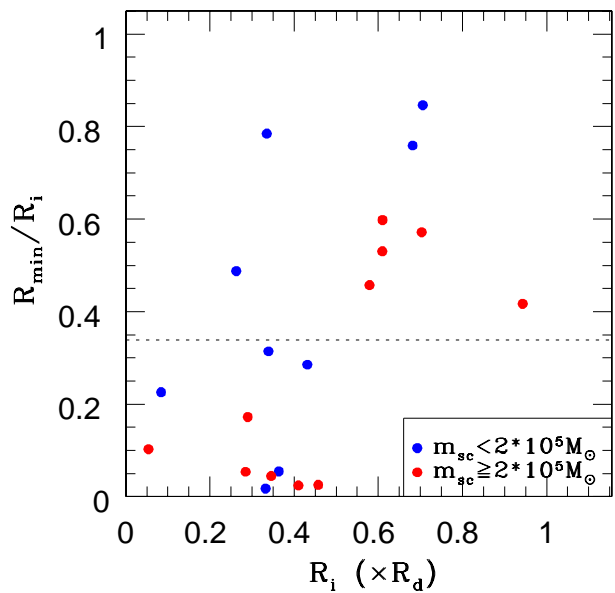,width=8.0cm}
\caption{
The ratios of minimum  $R$ ($R_{\rm min}$) to initial one ($R_{\rm i}$)
as a function of $R_{\rm i}$ for SCs in the standard model.
Blue and red dots represent SCs with 
$m_{\rm sc} <  2 \times 10^5 {\rm M}_{\odot}$
and $m_{\rm sc} \ge  2 \times 10^5 {\rm M}_{\odot}$, respectively.
The horizontal dotted line indicates the mean $R_{\rm min}/R_{\rm i}$
in this model.
}
\label{Figure. 5}
\end{figure}

\begin{figure}
\psfig{file=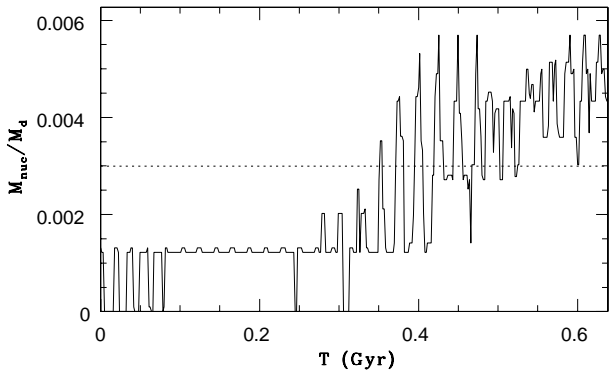,width=8.0cm}
\caption{
Time evolution of the mass-ratio of the nucleus 
to the disk ($M_{\rm nuc}/M_{\rm d}$)  in the standard model.
The dotted line is the observed mean of $M_{\rm nuc}/M_{\rm d}$.
See the main text for the details of the method  to
estimate $M_{\rm nuc}$.
}
\label{Figure. 6}
\end{figure}

\begin{figure}
\psfig{file=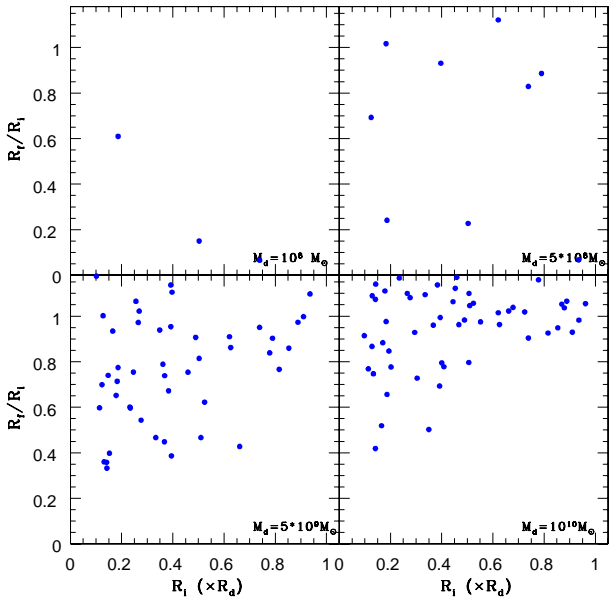,width=8.0cm}
\caption{
The ratios of final $R$ ($R_{\rm f}$) to initial one ($R_{\rm i}$)
as a function of $R_{\rm i}$ for SCs in four models with
$M_{\rm d}=10^{8} {\rm M}_{\odot}$ (upper left),
$M_{\rm d}=5 \times 10^{8} {\rm M}_{\odot}$ (upper right),
$M_{\rm d}=5 \times 10^{9} {\rm M}_{\odot}$ (lower left),
and $M_{\rm d}=10^{10} {\rm M}_{\odot}$ (lower right).
}
\label{Figure. 7}
\end{figure}

\begin{figure}
\psfig{file=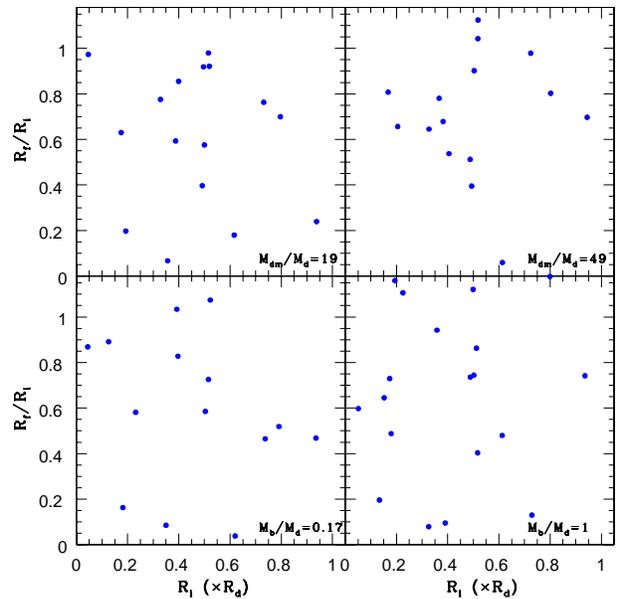,width=8.0cm}
\caption{
The same as Fig. 7.
but for four models with
$M_{\rm dm}/M_{\rm d}=19$ (upper left),
$M_{\rm dm}/M_{\rm d}=49$ (upper right),
$M_{\rm b}/M_{\rm d}=0.17$ (lower left),
and $M_{\rm b}/M_{\rm d}=1$ (lower right).
}
\label{Figure. 8}
\end{figure}

\begin{table*}
\centering
\begin{minipage}{185mm}
\caption{A brief summary of the results for the five representative
 models with different disk
masses.}
\begin{tabular}{ccccc}
{$M_{\rm d}$ ($\times 10^{9} M_{\odot}$)
\footnote{The total mass of a disk. For these five  disk models,
other model parameters such as $f_{\rm d}$ and $f_{\rm b}$ are exactly
the same between the five.}} &
{$M_{\rm nuc}$  ($\times 10^{6} M_{\odot}$)
\footnote{The total mass of the stellar nucleus in a disk: This is the total
mass of SCs within $0.1R_{\rm d}$. }} &
{$M_{\rm nuc}/M_{\rm d}$
\footnote{The nuclear mass fraction  of a disk. }} &
{$ \bar{ (\frac{ R_{\rm f} }{ R_{\rm i} } )} $ 
\footnote{The mean value
of $R_{\rm f}/R_{\rm i}$, where  $R_{\rm i}$ and $R_{\rm f}$ are
initial and final radii of SCs from the center of their host galaxy.
This  mean value  is estimated for SCs with $R_{\rm f}/R_{\rm i} \le 1$
so that we can 
more appropriately
measure the degree of orbital decay of SCs due to dynamical friction.}}  &
comments \\
0.1  & 0.5  &  0.0049  & 0.28 & low-mass disk model  \\
0.5  & 2.2  &  0.0043  & 0.50 &  \\
1.0  & 4.3  &  0.0043  & 0.55 & standard model  \\
5.0 &  3.0  &  0.0006  & 0.67 &  \\
10.0 & 2.8  &  0.00028  & 0.83 &   high-mass disk model \\
\end{tabular}
\end{minipage}
\end{table*}

\section{Results}
\subsection{The standard model}
 
Fig. 1 shows the time evolution of the projected distributions
of SCs for the last 0.64 Gyr 
in the standard model with $M_{\rm d}=10^{9} {\rm M}_{\odot}$
and with $N_{\rm sc}=20$.
The projected spatial distribution of SCs can change
with time not only because of the dynamical evolution of the disk
but also because of dynamical friction of SCs against the disk field
stars. The half-number radius of the SCS evolves from $0.38 R_{\rm d}$
(corresponding to $=0.86$ kpc)
to $0.30 R_{\rm d}$ ($=0.68$ kpc), which reflects that dynamical
friction of SCs can cause orbital decay of massive SCs within the disk.
The total number of SCs  within $0.1R_{\rm d}$  in the disk
can gradually increase so that a multiple SC system 
with the total mass ($m_{\rm sc}$)
of $4.3 \times 10^6 {\rm M}_{\odot}$
can form in the
nuclear region of the disk. 
The $x-$, $y-$, and $z-$components
of velocity dispersions of the multiple SCs are
17.7 km s$^{-1}$,  23.6  km s$^{-1}$, and 3.2  km s$^{-1}$. 
The velocity  dispersions comparable
to internal velocity dispersions of their nuclear  massive
SCs  suggest that the nuclear SCs can soon
merge to form a single nucleus.

Figs. 2 and 3 show orbital evolution projected onto the $x$-$y$ 
and $x$-$z$ planes
for SCs with two different masses.  In both case,
the distances from the center of the stellar disk ($R$) become
significantly smaller owing to orbital decay
caused by dynamical friction against disk field
stars. The extent to which a SC can decay its orbit owing to
dynamical friction depends on its initial mass and location
with respect to the center of its host galaxy:
the SCs with masses of $3.3 \times 10^5 {\rm M}_{\odot}$ and
$1.3 \times 10^6 {\rm M}_{\odot}$
can decrease $R$ by 31\% and 63\% of their initial $R$,
respectively.
The orbit of the more massive SC, which is initially in the inner
region of the disk, can be influenced also by the inner spirals and bar
developed during the disk evolution so that the orbital evolution can become
more chaotic. 
These orbital decay seen in these two SCs can be more clearly
seen in more massive SCs owing to more effective dynamical friction
for them (see the Appendix A for details of other SCs' orbits  in the standard
model). 

Fig.  4 shows that about 60 \% 
of the SCs have final $R$ ($R_{\rm f}$) significantly smaller than initial $R$
($R_{\rm i}$) owing to dynamical friction of the SC against disk field
stars. Spiral arms and bar in the disk can drive 
the angular momentum redistribution of the disk
so that $some$ SCs have  $R_{\rm f}<R_{\rm i}$: these are transferred  to
the outer regions of the disk by {\it global galactic dynamics}.
Fig. 4 also  shows that more massive SCs with $m_{\rm sc}
\ge  2 \times 10^5 {\rm M}_{\odot}$ are more likely to
have smaller $R_{\rm f}/R_{\rm i}$ (i.e., more significant
orbital decay due to dynamical friction).
The mean $R_{\rm f}/R_{\rm i}$ for the 20 SCs
at $T=0.64$ Gyr in this model is $\sim 0.85$.

Fig. 5 shows that all SCs can significantly decrease their $R_{\rm min}$
(corresponding to apo-center distances from the galaxy center)
owing to (i) dynamical friction and (ii) radial motions caused
by non-axisymmetric structures (bar and spiral arms).
Fig. 5 also shows that SCs with smaller $R_{\rm i}$ 
can experience  more significant orbital decay (i.e., smaller
$R_{\rm min}/R_{\rm i}$) due to dynamical friction,
because dynamical friction is more effective in higher densities
of background field stars.
The mean $R_{\rm min}/R_{\rm i}$ for the 20 SCs
at $T=0.64$ Gyr in this model is $\sim 0.34$.

Fig. 6 shows the time evolution of the total mass of SCs
(in units of $M_{\rm d}$)
within $0.1 \times R_{\rm d}$, which is referred to
as $M_{\rm nuc}$ for convenience. This $M_{\rm nuc}$ does not 
literally correspond to the mass of a single stellar
nucleus and thus goes up and down rapidly according to
time evolution of their orbits.
Clearly, around $T=0.4$ Gyr,
  $M_{\rm nuc}/M_{\rm d}$ can exceed  
the value of 0.003 observed for nucleated dwarf galaxies
in the Virgo cluster (C\^ot\'e  et al. 2006).
This relatively rapid formation of a multiple SC system
is due to very efficient dynamical friction of massive SCs
with the mean mass of  $1.1  \times 10^6 {\rm M}_{\odot}$
in the disk. Therefore, the time evolution of  $M_{\rm nuc}/M_{\rm d}$
depends strongly on the adopted initial mass function of
SCs (this point is discussed later).

\subsection{Parameter dependences}
Dynamical friction processes of SCs in disk galaxies 
depend strongly  on model parameters
such as $M_{\rm d}$, $M_{\rm dm}/M_{\rm d}$ ($=1/f_{\rm d}-1$),
 $M_{\rm b}/M_{\rm d}$
(=$f_{\rm b}$), central stellar densities of disks,
and $M_{\rm V, high}$.
The derived $M_{\rm d}$-dependence is the most important result
in the present study
so that it is briefly summarized in the Table  2. 
We illustrate  the derived dependences on model parameters as follows:

(i) Orbital decay of SCs due to dynamical friction
is more important in less massive host galaxies, mainly
because relative velocities ($v_{\rm rel}$) between SCs and disk field stars
(which are comparable to stellar velocity dispersions of the disks)
are smaller in the galaxies: the frictional drag force can be 
proportional to $v_{\rm rel}$ (see 7.18 in Binney \& Tremaine 1987).
Fig. 7 shows that $R_{\rm f}/R_{\rm i}$ is systematically smaller
in disk galaxies with smaller  $M_{\rm d}$ for a fixed $f_{\rm d}$.
Fig. 7 also shows that for disks with $M_{\rm d} \ge 5 \times
10^{10} {\rm M}_{\odot}$,  no/few SCs can show significant
orbital decay ($R_{\rm f}/R_{\rm i}<0.4$), which means that
dynamical friction of SCs is not so important in their orbital evolution.

(ii) Dynamical friction of SCs by disk field stars is less
effective in galaxies with smaller $f_{\rm d}$, mainly
because stellar densities of disk field stars are lower
so that frictional drag force can be weaker.
Fig. 8  shows that for disks with $M_{\rm d}/M_{\rm d}=49$,
only a few SCs can show significant
orbital decay ($R_{\rm f}/R_{\rm i}<0.4$), which implies  that
dynamical friction of SCs can not be  so important in
orbital evolution of SCs in the early phase of disk evolution
of galaxies when the mass fractions of disk stars can be
very small.
Fig. 8 also shows that a smaller fraction of SCs have 
$R_{\rm f}/R_{\rm i}>1$, which reflects that radial redistribution
of SCs due to the presence of non-axisymmetric structures
(spirals and bars) does not happen in galaxies
with smaller degrees of self-gravitating.

(iii) Galactic bulges appear to be less important in orbital
decay of SCs within disks in comparison with other galactic
properties such as $M_{\rm d}$ and $f_{\rm d}$.
Fig. 8 shows  no remarkable  differences between models
with different $f_{\rm b}$ (=0.17 and 1.0), though
$f_{\rm b}$ can be an important parameter which determines
whether stellar bars can form and thus affect
orbital evolution of SCs.

(iv) The time scale of dynamical friction of SCs by disk field stars
is longer in LSBs than in HSBs,  because stellar densities 
of background field stars can be  
key factors for the effectiveness of dynamical friction.
As a result of this,  formation of stellar nuclei via SC merging
can proceed more slowly in LSBs. For example,
$M_{\rm nuc}$ can become  $\sim 10^6 {\rm M}_{\odot}$ 
only after 2.4Gyr orbital evolution
of SCs in the LSB model in which $R_{\rm d}$ is by a factor of 2.5
larger than that in the standard model (and other model parameters
are the same as those in the standard model).

(v) Final properties of SCSs (e.g., $R_{\rm h,sc}$) 
and the formation processes of nuclear multiple SCs (thus
nucleus formation processes) in galaxies
depend on initial MFs (e.g., $M_{\rm V,high}$).
For example, $R_{\rm h,sc}$ can evolve more rapidly 
and have smaller final values
in models with smaller  $M_{\rm V,high}$ (i.e., larger
$m_{\rm sc}$ for the most massive SCs) in the present
study (e.g.,  $R_{\rm h,sc}=0.2R_{\rm d}$
for models with $M_{\rm V,high}=-11$ mag).
Also final $M_{\rm nuc}$ is larger in models
with smaller $M_{\rm V,min}$ 
(e.g., $M_{\rm h, nuc}/M_{\rm d}=0.02$ 
for $M_{\rm V, high}=-11$ mag).
This result implies that formation processes of stellar galactic
nuclei (e.g., whether dwarfs are nucleated or non-nucleated)
depend on original MFs of SCs in  galaxies.

\begin{figure}
\psfig{file=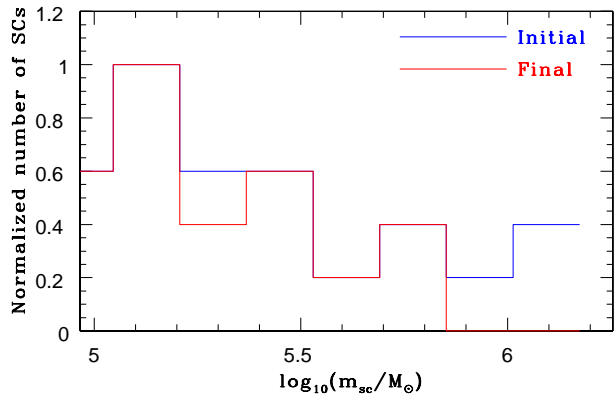,width=8.0cm}
\caption{
The initial (blue) and final (red) distributions of $m_{\rm sc}$
in the standard model. The details of the method  to derive
these SCMFs are given in the main text.
}
\label{Figure. 9}
\end{figure}

\begin{figure}
\psfig{file=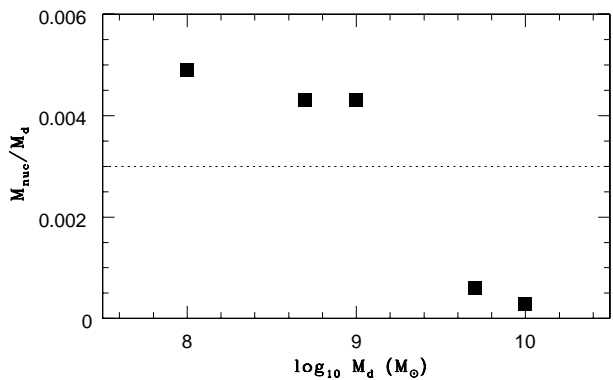,width=8.0cm}
\caption{
The dependence of $M_{\rm nuc}/M_{\rm d}$ on
$M_{\rm d}$ for five models with different $M_{\rm d}$.
The  dotted horizontal line indicates the observed value ($=0.003$).
}
\label{Figure. 10}
\end{figure}

\begin{figure}
\psfig{file=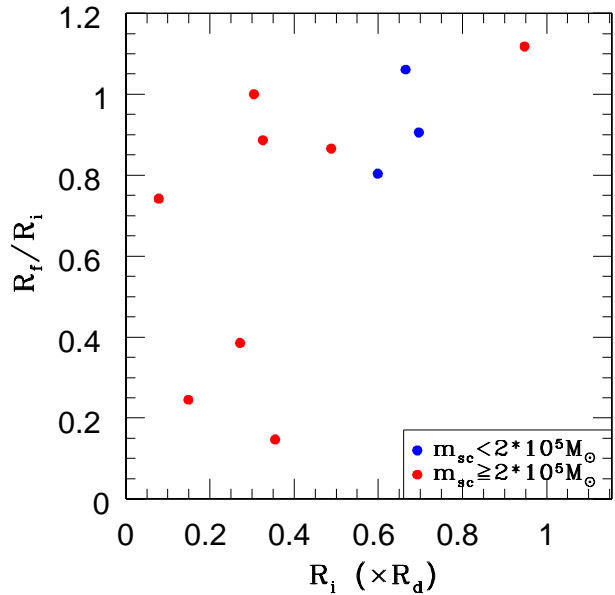,width=8.0cm}
\caption{
The same as Fig. 4 but for halo SCs in the standard model.
Since 9 out of 20 have $R_{\rm f}/R_{\rm i}>1.2$,
they can not be seen in the plot. 
Since $R_{\rm i}$ of a halo SC is not set to be the pericenter distance
of its orbit initially, $R_{\rm f}$ can be significantly
larger than $R_{\rm i}$ during its orbital evolution. 
The  mean value of $R_{\rm f}/R_{\rm i}$
for all SCs is 1.5 and that for SCs with $R_{\rm f}/R_{\rm i} \le 1$
is 0.64
in this model.
These   suggest  that dynamical
friction of halo SCs is less effective than that of disk SCs
in disk galaxies.
}
\label{Figure. 11}
\end{figure}

\section{Discussion}

\subsection{Evolution of SC luminosity functions}

The present study has first shown that more massive SCs can spiral-in to
the nuclear regions of galactic disks owing to dynamical friction
within less than $\sim 1$ Gyr in low-mass disk galaxies.
It is highly  likely that these multiple SCs in
nuclear regions of disk galaxies can merge with
one another to form stellar galactic nuclei (or single massive nuclear SCs).
If this is the case,
then  the merged SCs (i.e., stellar galactic nuclei or nuclear SCs) 
would not be identified  as ordinary SCs in observations.
Therefore, it is meaningful for the present study to investigate SCMFs
located within $R\ge0.1R_{\rm d}$ for comparing the simulated SCMFs
with the observed ones. In the followings, we assume that
MFs are for SCs located within $R\ge0.1R_{\rm d}$  (i.e., excluding
the nuclear SCs in the simulations).

Fig. 9 shows that no SCs with 
${\log}_{10}(m_{\rm sc}/{\rm M}_{\odot}) > 5.9$
can be seen in the final MF for the standard model, because 
more massive SCs spiral-in to nuclear regions
of disk galaxies more rapidly owing to dynamical
friction of SCs against  disk field stars.
This result implies that MFs of SCs for 
$5.0 <  {\log}_{10}(m_{\rm sc}/{\rm M}_{\odot})$ (i.e., higher-mass end)
can steepen significantly in comparison of their original MFs (at the epoch
of SC formation) owing to dynamical friction of SCs.
Furthermore, given that dynamical friction of SCs against  disk field stars
is more effective  in less massive disk galaxies,
the MFs can more significantly steepen in less massive disk galaxies.
Thus, it is highly likely that the SCMFs can be steeper in
less massive disk galaxies.

Based on observational studies of GCLFs for the early-type galaxies
in the ACS Virgo Cluster Survey,
Jord\'an et al. (2006) showed that the GCMFs for relatively
high masses ($m_{\rm sc} \ge 3 \times 10^5 {\rm M}_{\odot}$) are steeper
in less luminous galaxies in the Virgo cluster.
They suggested that since GCMFs are not strongly affected by dynamical
evolution of GCs (e.g., dynamical friction) over a Hubble time,
the steeper GCMFs arise from variations in initial conditions.
Our results however imply that dynamical friction of SCs against disk field
stars is so effective as to change the shapes of GCMFs within a relatively
short time scale.  Our results also imply that if these less luminous
early-type galaxies were formed from disk galaxies either by
galaxy interactions or by major mergers, their GCMFs can be steeper
owing to changes of  GCMFs in {\it their progenitor disks}
 caused by dynamical friction.

\subsection{Formation of stellar galactic nuclei}

Since Tremaine et al. (1975) first investigated dynamical evolution
of GCs interacting with background stars through dynamical friction
in M31 to discuss nucleus formation in M31,
many authors discussed formation of stellar galactic nuclei via
mergers of SCs (or GCs) in galaxies
with different physical properties 
(e.g., Lotz et al. 2001; Capuzzo-Dolcetta \& Vicari 2005).
Recent fully  self-consistent N-body simulations have 
shown that merging of SCs in the inner regions of galaxies
lead to the formation of stellar galactic nuclei with their
projected radial density profiles similar to the observed ones 
(e.g., Capuzzo-Dolcetta \& Miocchi  2008).
Although these works did not discuss potential importance
of dissipative gas dynamics and star formation in the formation
of stellar galactic nuclei (e.g., Bekki et al. 2006; Bekki 2007),
they significantly advanced our understanding dynamical evolution
of multiple SC systems in galaxies.

One of remaining questions in the SC merger scenario of nucleus formation
is whether the scenario can explain the observed relation between
stellar nucleus masses  and their host galaxy ones 
($M_{\rm nuc} \sim 0.003 M_{\rm g}$,
where $M_{\rm g}$ is the galaxy mass; C\^ot\'e et al. 2006).      
Since the observational study  by   C\^ot\'e et al. (2006)  derived 
total mass of the luminous components 
(i.e., not  those for entire galaxies including
the extended dark matter halos) of galaxies by using reasonable $M/L$,
it is reasonable  to use $M_{\rm d}$ rather than $M_{\rm gal}$ for
the purpose of discussing the observed $M_{\rm nuc}=0.003 M_{\rm g}$ relation.
Fig.  10 shows $M_{\rm nuc}/M_{\rm d}$ as a function of $M_{\rm d}$
for the five models in which only  $M_{\rm d}$  are different between them.
The simulated $M_{\rm nuc}/M_{\rm d}$ is close to the observed value of 0.003
for less luminous disk galaxies with $M_{\rm d} \le 10^9 {\rm M}_{\odot}$.
SCs can lose a significant fraction of their original
masses during their orbital evolution  in galaxies 
(e.g., Fujii et al. 2008), which is not considered
at all in estimating $M_{\rm nuc}/M_{\rm d}$ in the present study.
Thus the simulated $M_{\rm nuc}/M_{\rm d}$ for less luminous disks
can be regarded as fairly consistent with observations.

Fig. 10 also shows that formation of stellar galactic nuclei with 
$M_{\rm nuc}/M_{\rm d} \sim 0.003$ by SC merging is not possible
in more luminous disks owing to ineffective dynamical friction.
Therefore, the result implies that if the observed  $M_{\rm nuc}=0.003M_{\rm g}$
relation can be hold for all galaxies,  the SC merger scenario may well
have a serious problem in explaining the observation.
However, 
galaxies are more likely to have massive black holes (MBHs) rather
than stellar nuclei in more luminous galaxies with $M_{\rm g} \ge 10^{10} {\rm M}_{\odot}$ 
(e.g., Ferrarese et al. 2006),
 though it is not so clear whether  these galaxies with MBHs 
also have stellar nuclei (i.e., MBHs and stellar nuclei coexist in galaxies).
Thus it would be currently reasonable to claim that the SC merger scenario is a promising
one for the formation of stellar galactic nuclei in galaxies.

\subsection{Compassion with dynamical friction by galactic halos}

We have first shown that dynamical friction of SCs against disk 
field stars can be so effective in low-mass disk galaxies
so that orbital decay of SCs can be as significant as to change
physical properties (e.g., half-number radii and SCMFs)
of their SCSs.
Previous theoretical and numerical works focused exclusively
on dynamical friction of halo SCs 
(e.g., old GCs) against 
dark matter halos and 
showed that the time scale of orbital decay of SCs due
to dynamical friction against dark matter halos is relatively
long (a few to several Gyr) even for low-mass dwarfs like
the Fornax (e.g., Goerdt et al. 2006 for the cuspy NFW profile
of its dark matter halo).
Since they did not investigate the influences of stellar disks
on orbital evolution of halo SCs through dynamical friction,
it is worthwhile for the present study to investigate
how halo SCs  evolve in disk galaxies if they can experience
dynamical friction against both dark matter halos and
against  disk field stars.

Fig. 11 shows that orbital decay of halo SCs due to dynamical friction
against halos and disks is much less significant
owing to (i) lower mass densities of the halos and (ii) larger velocity differences
in the SCs and dark matter halos for the standard model.
The mean $R_{\rm f}/R_{\rm i}$ for the halo SCs is 1.53,
which is by a factor of $\sim 2$  larger than that for the disk SCs.
Only some more massive SCs with $m_{\rm sc} \ge 2 \times 10^5 {\rm
M}_{\odot}$ can show significant orbital decay (i.e., decreasing
apocenter distances of their orbits) for halo SCs.
This less effective dynamical friction in halo SCs is a common
feature in the present models,
which suggests that the presence of disks in galaxies
does not significantly change dynamical friction processes
of halo SCs.  It should be however stressed
that the halo SCs initially in the inner region of the disk galaxy
in the standard model
shows a slightly flattened final spatial distribution: this can be due to
relatively stong dynamical friction between halo SCs and disk field stars
in the inner regions of galaxies.

\section{Conclusions}

We have investigated orbital  evolution of SCs being influenced by
dynamical friction of disk field stars
based on galaxy-scale numerical simulations
on dynamical evolution of disk galaxies.
We have investigated models with
variously different model parameters in order to understand
dependences of the effectiveness of dynamical friction of SCs 
on physical properties of disk galaxies.
We summarize our
principle results as
follows.

(1) Dynamical friction of SCs against  disk field stars is much more effective
in orbital decay of SCs in comparison with that against  galactic halos
in disk galaxies.
For example, the half-number radius of the SCS  ($R_{\rm h, sc}$) in a galaxy with
$M_{\rm d}=10^9 {\rm M}_{\odot}$ can decrease by $\sim 30$\%
owing to orbital decay of SCs caused by dynamical friction  within
well less than $10^9$ yr for a reasonable MF of SCs.
However orbital decay of SCs due to dynamical friction can be 
more clearly seen 
only for more massive SCs (e.g., those with $m_{\rm sc} \ge 2 \times 10^5
{\rm M}_{\odot}$).

(2) Dynamical friction processes of SCs against disk field stars depend on
physical properties of their host galaxies.  For example,
the dynamical friction is much more effective
in disks with lower $M_{\rm d}$ owing to smaller stellar velocity dispersions.
As a result of this, $R_{\rm h, sc}$ can more rapidly evolve with time
in disks with lower  $M_{\rm d}$. 
Evolution of SCSs in more massive disks 
(with $M_{\rm d} \sim 10^{10} {\rm M}_{\odot}$)
can not be clearly seen in the present study, which implies
dynamical friction is important only for 
evolution of  SCSs (e.g., $R_{\rm h, sc}$)
in less massive disk galaxies.

(3) Dynamical friction of  SCs by disk field stars is more effective
in disks with higher degrees of self-gravitating of the disks 
(e.g., higher $f_{\rm d}$), which implies  that higher
stellar densities and non-axisymmetric structures
such as bars and spiral arms can  promote dynamical friction.
Orbital decay of SCs due to dynamical friction
in disks does not depend so strongly on bulge masses of the disks
for $0 \le f_{\rm b} \le 1$ in the present study.
Evolution of SCSs due to orbital decay of SCs caused
by dynamical friction is more important  in HSBs than in LSBs.

(4) More massive SCs can spiral-in to the nuclear regions ($R<0.1R_{\rm d}$)
of disk galaxies owing to dynamical friction so that they can form multiple SC systems there
in less massive disk galaxies with $M_{\rm d} \le  10^{9} {\rm M}_{\odot}$.
These systems  show velocity dispersions comparable to or less than
internal stellar velocity dispersions of SCs so that they are likely to
merge quickly with one another to form single massive SCs there.
Thus formation of stellar galactic nuclei via merging of more massive SCs
is possible in less massive disk galaxies.

(5) The observed $M_{\rm nuc}-M_{\rm g}$ relation 
(i.e., $M_{\rm nuc} \sim  0.003 M_{\rm g}$) for less luminous
galaxies can be closely associated with formation of stellar galactic
nuclei by SC merging.  The LF/MFs  for more massive  SCs
in SCSs of disk galaxies can steepen
owing to the formation of stellar galactic nuclei through
merging of more massive SCs inwardly transferred to the nuclear region
of galaxies.
This steeping of SCLF/MFs due to dynamical friction
can be more significant in less massive  galaxies.

\section{Acknowledgment}
I am   grateful to the anonymous referee for valuable comments,
which contribute to improve the present paper.
KB acknowledges the financial support of the Australian Research
Council
throughout the course of this work.
Numerical computations reported here were carried out both on the GRAPE
system
at the University of New South Wales  and on those  
kindly made available by the Center for computational
astrophysics
(CfCA) of the National Astronomical Observatory of Japan.

\appendix

\section{Orbital evolution of individual SCs}

\begin{figure}
\psfig{file=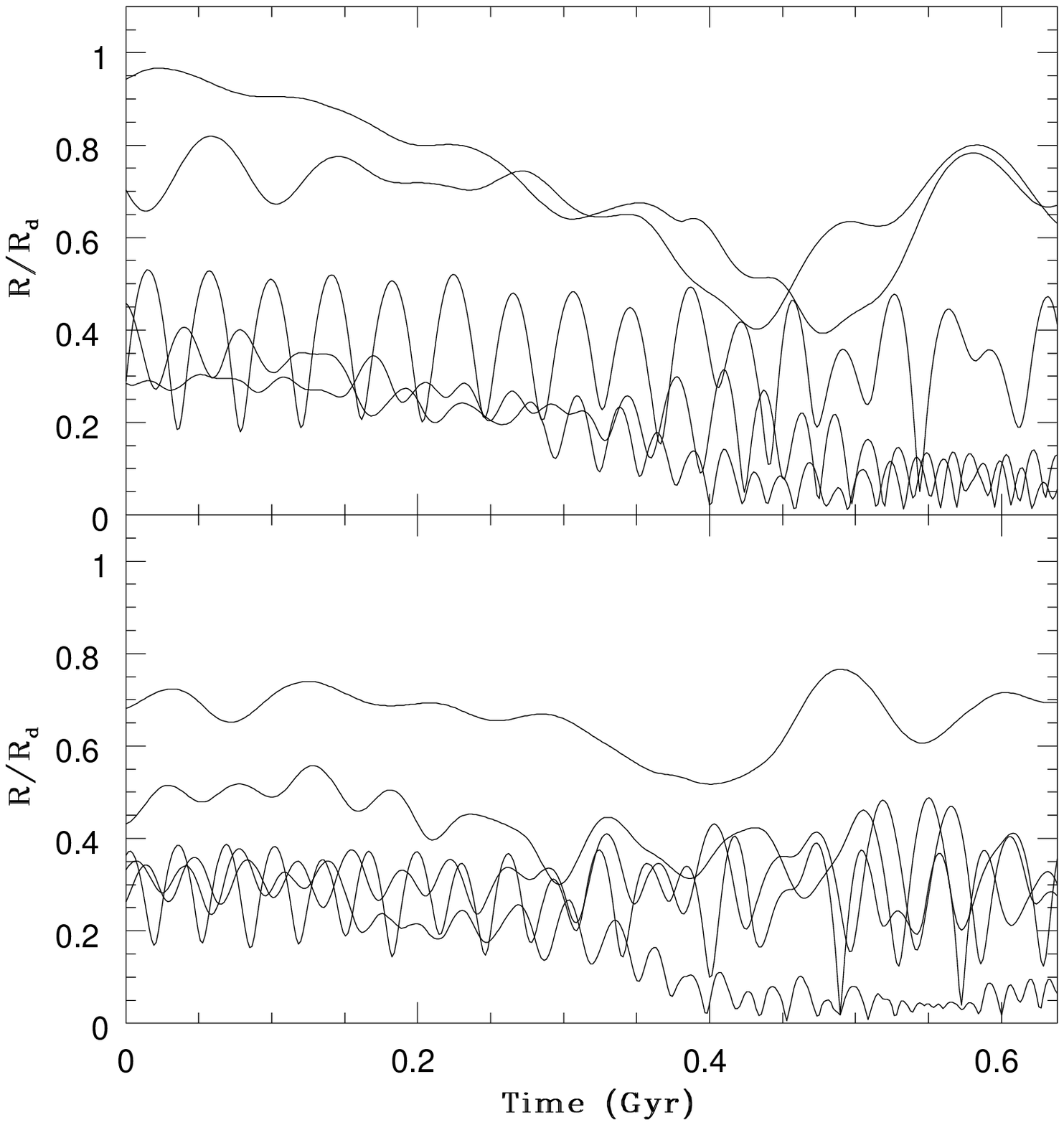,width=8.0cm}
\caption{
Time evolution of $R/R_{\rm d}$ of five SCs selected
from those with $m_{\rm sc}<2 \times 10^5 {\rm M}_{\odot}$
(lower) and  from 
those with $m_{\rm sc} \ge 2 \times 10^5 {\rm M}_{\odot}$ (upper)
in the standard model. Owing to dynamical infleunces of 
a stellar bar and spiral arms on SCs and SC-SC mutual
interaction,    $R/R_{\rm d}$ of these SCs
can evolve in a complicated way.
}
\label{Figure. A1}
\end{figure}

Fig. A1 shows time evolution of distances ($R$
in units of $R_{\rm d}$) of ten SCs from
the center of their host disk galaxy in the standard model.
In order to show more clearly orbital evolution of each individual SC,
only 5 SCs are selected and shown for less massive
SCs ($m_{\rm sc} < 2 \times 10^5 {\rm M}_{\odot}$)
and for more massive ones ($m_{\rm sc} \ge 2 \times 10^5 {\rm M}_{\odot}$).
The orbital eccentricity  of a SC in the present study is described as
follows:
\begin{equation}
e_{\rm orb}= \frac{ R_{\rm max}-R_{\rm min} }{ R_{\rm max}+R_{\rm
min} },
\end{equation}
where $R_{\rm min}$ and $R_{\rm max}$ are the minimum and maximum
$R$ for the SC.
It is found that $R_{\rm min}$  of the selected ten SCs range from 0.006 to
0.51 and  
the mean $R_{\rm min}$ is 0.17.
It is also found that $e_{\rm orb}$  of the SCs range from 0.19 to
0.97 and  
the mean $e_{\rm orb}$ is 0.67.

Owing to dynamical interaction between  the non-axisymmetric
structures (e.g., bar and spiral arms)
developed in the later phase of  the disk's 
dynamical  evolution,
orbital evolution of these SCs can be quite complicated.
One example of this is that
not all of the SCs show significant declines in $R$ 
owing to dynamical friction
within $\sim 0.6$ Gyr.
Some of SCs show orbital decay before the formation
of a stellar bar, their $R$ can again become larger
after the bar formation
owing to redistribution of angular momentum by the bar.
It appears that SC-SC interaction does not play a vital
role in the orbital evolution of SCs owing to the small
number ($=20$) of SCs in this model.

\end{document}